\documentclass[letterpaper,10pt,conference]{ieeeconf}

\IEEEoverridecommandlockouts
\usepackage{amsmath,amssymb,bm}
\usepackage{graphicx}
\usepackage{booktabs}
\usepackage{multirow}
\usepackage{placeins}
\usepackage{float}

\title{\LARGE \bf
VISKY: Virtual Inertia Skyhook Control for Semi-Active Suspension Systems Using Magnetorheological Dampers
}

\author{Hansol Lim$^{1,2}$, Jee Won Lee$^{1,2}$, Seung-Bok Choi$^{1,3*}$, and Jongseong Brad Choi$^{1,2*}$%
\thanks{$^{1}$Department of Mechanical Engineering, State University of New York, Korea, 21985, Republic of Korea.}%
\thanks{$^{2}$Department of Mechanical Engineering, State University of New York, Stony Brook, Stony Brook, NY 11794, USA.}%
\thanks{$^{3}$Department of Mechanical Engineering, Industrial University of Ho Chi Minh City, Ho Chi Minh City, 727000, Vietnam.}%
\thanks{E-mails: hansol.lim@stonybrook.edu; jeewon.lee@stonybrook.edu; seungbok.choi@stonybrook.edu; jongseong.choi@stonybrook.edu.}%
\thanks{This work is supported by the National Research Foundation of Korea (NRF) grant funded by the Korea government (MSIT) (Grant No. RS-2022-NR067080 and RS-2025-05515607). *Corresponding authors: Seung-Bok Choi, Jongseong Brad Choi.}%
}

\begin{document}
\maketitle
\thispagestyle{empty}
\pagestyle{empty}

\begin{abstract}
This paper presents a Virtual Inertia Skyhook (VISKY) controller for magnetorheological (MR) dampers in semi-active suspensions. The proposed law is derived from a continuous sky-ground damping baseline augmented with acceleration feedback on the sprung and unsprung masses. In the closed-loop equations, these acceleration terms appear as a mass-like virtual inertia matrix rather than as a change in physical hardware. This interpretation motivates the VISKY name while making the underlying sky-ground hybrid structure explicit. Numerical evaluations under half-sine bump, representative ISO 8608 random-road, and stepped sine-sweep inputs show reduced body-acceleration and wheel-acceleration metrics relative to conventional Skygroundhook, with the largest gains appearing near the wheel-hop mode. The controller retains low computational overhead because it requires only algebraic force computation and bounded MR inversion.
\end{abstract}

\section{Introduction}
Semi-active suspension systems are increasingly used for a variety of ground vehicles, from passenger cars to mobile robots \cite{c1,c2}. They offer a good balance between passive and fully active suspensions to achieve adaptive performance with comparatively minimal power requirements \cite{c3}. These systems can respond quickly to irregular terrain and other disturbances by reducing vibrations experienced by the vehicle body. A common actuator for modern semi-active suspensions is the magnetorheological (MR) damper. It adjusts its viscosity based on an applied magnetic field \cite{c4,c5}. With a very fast response time and relatively low energy demand for its damping force, MR dampers have been successfully demonstrated in commercial vehicles and robotic platforms \cite{c6}.

One of the most widely used semi-active control strategies for wheeled vehicles and robots is the Skyhook principle. It aims to reduce vertical body acceleration by having a virtual damper connected to a fixed inertial reference in the sky \cite{c7}. This approach significantly improves ride comfort by prioritizing the reduction of vibrations felt by the vehicle body. On the other hand, Groundhook control is a variant of Skyhook control. It focuses on reducing vibration felt on the wheels instead of the body \cite{c8}. Although each method excels at its objective, neither is optimal for both ride comfort and road holding under all conditions. Consequently, a hybrid of the two called the Skygroundhook controller was introduced. It blends the two approaches to reduce both body and wheel oscillations simultaneously \cite{c9,c10,c11}. While Skygroundhook yields improved overall performance, it has notable weaknesses:

\begin{enumerate}
\item Standard Skygroundhook often addresses a narrow frequency band effectively but can transmit higher vibrations at off-nominal frequencies \cite{c12}.
\item When transitioning between Skyhook and Groundhook phases, the system might experience abrupt force changes, introducing unwanted high-frequency vibrations that may resonate with higher modes \cite{c13}.
\end{enumerate}

These challenges motivate more advanced controllers that harness the flexibility of MR dampers. Over the past decade, many innovative strategies were published, from adaptive controllers to AI-driven methods such as fuzzy logic \cite{c9,c10,c14} and neural networks \cite{c15,c16}, which have aimed to refine semi-active suspension performance.

However, these algorithms can be computationally demanding, requiring powerful microprocessors or specialized hardware. Not all automotive or robotic suspensions can accommodate such costs or complexities. Consumer vehicles and small mobile robots have limited onboard processing capacity and may not handle these computationally demanding techniques.

Therefore, we propose a Virtual Inertia Skyhook (VISKY) controller for MR dampers. Although the acronym is retained, the controller is not a pure Skyhook law. Instead, it extends a continuous Skygroundhook baseline with additional acceleration feedback on both the sprung and unsprung sides. In the resulting closed-loop equations, these acceleration terms form a mass-like virtual inertia matrix that improves high-frequency attenuation without adding physical hardware. The MR damper then tracks the commanded semi-active force with bounded voltage input \cite{c17}.

The remainder of this paper is structured as follows. Section II presents the background and problem formulation. Section III derives the VISKY control law and MR damper realization. Section IV provides the linear stability result. Section V introduces gain tuning. Section VI presents simulation-based results on bump, random-road, and stepped sine-sweep excitation. Section VII concludes the paper.

Accordingly, this paper makes three contributions. First, it formulates a sky-ground hybrid semi-active control law with additional acceleration feedback and shows how the added terms induce a virtual inertia matrix in the closed-loop quarter-car model. Second, it derives an MR damper realization with explicit bounded-input handling and states a linear stability condition for the ideal force-tracking model. Third, it evaluates VISKY against tuned Skyhook, Groundhook, and Skygroundhook baselines using bump, ISO 8608 random-road, and stepped sine-sweep simulations.

\section{Background and Problem Formulation}
\subsection{Quarter-Car Model}
Semi-active suspensions are often analyzed through a quarter-car model, where a sprung mass $m_s$ (vehicle body) is connected to an unsprung mass $m_u$ (wheel assembly) via a spring $k_s$ and a passive damper $c_s$. The wheel contacts the ground through a tire modeled by $k_t$. Let $z_s$ and $z_u$ be the vertical displacements of the sprung and unsprung masses, respectively. A road input $z_r$ represents terrain elevation. The equations of motion are
\begin{equation}
 m_s\ddot z_s = -k_s(z_s-z_u)-c_s(\dot z_s-\dot z_u)+F
 \tag{1a}
\end{equation}
\begin{equation}
 m_u\ddot z_u = k_s(z_s-z_u)+c_s(\dot z_s-\dot z_u)-k_t(z_u-z_r)-F.
\tag{1b}
\end{equation}

\subsection{Classical Skyhook, Groundhook, and Skygroundhook Control}
The Skyhook control is a classic method that reduces sprung-mass vibration by emulating a virtual damper between $z_s$ and an inertial sky. A common on-off version is
\begin{equation}
F_{\mathrm{sky}}=
\begin{cases}
-C_{\mathrm{sky}}\dot z_{\mathrm{rel}}, & \dot z_s\dot z_{\mathrm{rel}}>0,\\
0, & \dot z_s\dot z_{\mathrm{rel}}\le 0.
\end{cases}
\tag{2}
\end{equation}
where $\dot z_{\mathrm{rel}}=\dot z_s-\dot z_u$.

Groundhook control focuses on unsprung-mass stability:
\begin{equation}
F_{\mathrm{gr}}=
\begin{cases}
 -C_{\mathrm{gr}}\dot z_{\mathrm{rel}}, & \dot z_u\,\dot z_{\mathrm{rel}}<0,\\
0, & \dot z_u\,\dot z_{\mathrm{rel}}\ge 0.
\end{cases}
\tag{3}
\end{equation}

The Skygroundhook method combines both strategies:
\begin{equation}
F_{\mathrm{sg}}=
\begin{cases}
 -C_{\mathrm{sky}}\dot z_s-C_{\mathrm{gr}}\dot z_u, & \dot z_s\dot z_{\mathrm{rel}}>0,\\
 -C_{\mathrm{passive}}\dot z_{\mathrm{rel}}, & \dot z_s\dot z_{\mathrm{rel}}\le 0.
\end{cases}
\tag{4}
\end{equation}
with
\begin{equation}
C_{\mathrm{eq}}=\mathrm{clamp}\!\left(-\frac{F_{\mathrm{sg}}}{\dot z_{\mathrm{rel}}},0,C_{\max}\right).
\tag{4a}
\end{equation}
where $F_{\mathrm{sg}}$ is the hybrid Skygroundhook force in (4), $C_{\mathrm{passive}}$ is the passive damping coefficient used in the fallback branch, $C_{\mathrm{eq}}$ is the equivalent damping command, and $\mathrm{clamp}(x,0,C_{\max})=\min(\max(x,0),C_{\max})$.
Although hybrid controllers yield better overall performance, they still exhibit frequency-dependent limitations, abrupt switching, and incomplete adaptability under variable terrain.

\subsection{MR Damper Model}
Magnetorheological fluid comprises micron-scale ferrous particles in a carrier fluid. Under an applied magnetic field, these particles form chain-like structures that increase fluid yield stress, thus varying effective damping coefficient \cite{c9,c10}. A typical MR damper can switch from near-minimal damping to high damping within milliseconds by altering the voltage command $u$. One widely adopted MR damper model is the Bouc-Wen formulation \cite{c18}:
\begin{equation}
F_{\mathrm{MR}}=(C_{0a}+C_{0b}u)\dot z_{\mathrm{rel}}+k_0 z_{\mathrm{rel}}+(\alpha_a+\alpha_bu)\xi
\tag{5}
\end{equation}
\begin{equation}
\dot \xi=-q|\dot z_{\mathrm{rel}}|\xi-b\dot z_{\mathrm{rel}}|\xi|+\gamma\dot z_{\mathrm{rel}}.
\tag{5a}
\end{equation}
Here, $z_{\mathrm{rel}}=z_s-z_u$, $\dot z_{\mathrm{rel}}=\dot z_s-\dot z_u$, and $\xi$ is the Bouc-Wen hysteresis state.

\subsection{Related Derivative-Based Semi-Active Control}
Beyond classical Skyhook/Groundhook/hybrid methods, many adaptive and intelligent controllers use MR damper tunability more effectively \cite{c9,c10,c19}. Fuzzy logic controllers handle nonlinearities with rule-based inference, and neural networks learn complex state-to-voltage mappings \cite{c15}. These methods can be computationally expensive in cost-sensitive systems \cite{c20}.

A tuned mass damper (TMD) is a well-established hardware approach to vibration control, but it adds mass and packaging complexity and loses performance away from its tuned frequency. To overcome these issues, researchers have explored efficient enhancements to classical logic using derivative-based feedback. Tyan \cite{c21} used an $H_\infty$ controller with proportional-derivative action for robustness, and Duan \cite{c22} applied state-derivative feedback to structural control. These studies support derivative-based damping augmentation, but they generally target a single objective and do not fully integrate with Skygroundhook hybrid logic.

Our proposed VISKY controller addresses this gap by extending continuous Skygroundhook with acceleration feedback, thereby introducing a virtual inertia matrix into the closed-loop model without physical TMD complexity or AI-level computational burden.

\section{VISKY Control Design}
This section derives VISKY from a continuous Skygroundhook baseline augmented with velocity-acceleration feedback and then integrates it with the MR damper model. Although we retain the acronym VISKY, the controller is not a pure Skyhook law: it contains both sprung-side and unsprung-side terms, so it should be interpreted as a sky-ground hybrid with a virtual inertia matrix. Stability is analyzed in Section IV.
\begin{figure}[!t]
\centering
\includegraphics[width=0.58\columnwidth]{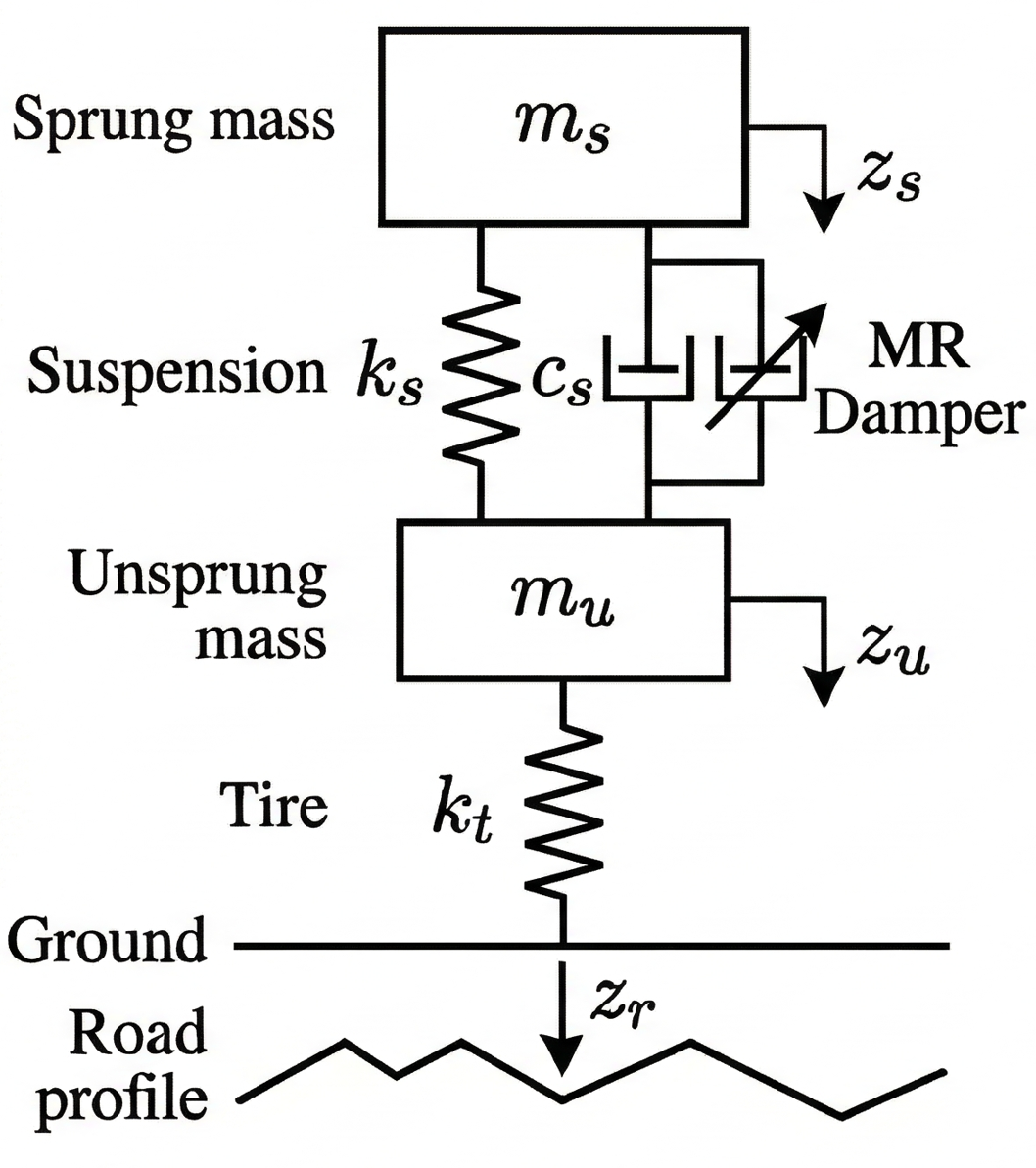}
\caption{Quarter-car model with MR damper integration.}
\label{fig:qcar_schematic}
\end{figure}

\subsection{VISKY Control Law}
The desired damping force is defined as
\begin{equation}
F_d=-P_{\mathrm{sky}}\dot z_s-D_{\mathrm{sky}}\ddot z_s-P_{\mathrm{gr}}\dot z_u-D_{\mathrm{gr}}\ddot z_u.
\tag{6}
\end{equation}
Here, $P_{\mathrm{sky}}$ and $P_{\mathrm{gr}}$ have damping-like units of $\mathrm{N\cdot s/m}$, whereas $D_{\mathrm{sky}}$ and $D_{\mathrm{gr}}$ have mass-equivalent units of $\mathrm{kg}$. When $D_{\mathrm{sky}}=D_{\mathrm{gr}}=0$, (6) reduces to a continuous sky-ground damping baseline. When $P_{\mathrm{gr}}=D_{\mathrm{gr}}=0$, it becomes a skyhook-like variant. When $P_{\mathrm{sky}}=D_{\mathrm{sky}}=0$, it becomes a groundhook-like variant.

The closed-loop quarter-car dynamics become
\begin{equation}
m_s\ddot z_s=-k_s(z_s-z_u)-c_s(\dot z_s-\dot z_u)+F_d
\tag{7}
\end{equation}
\begin{equation}
m_u\ddot z_u=k_s(z_s-z_u)+c_s(\dot z_s-\dot z_u)-k_t(z_u-z_r)-F_d.
\tag{8}
\end{equation}
Collecting acceleration terms yields
\begin{equation}
(m_s+D_{\mathrm{sky}})\ddot z_s + D_{\mathrm{gr}}\ddot z_u = b_1
\tag{9}
\end{equation}
\begin{equation}
-D_{\mathrm{sky}}\ddot z_s + (m_u-D_{\mathrm{gr}})\ddot z_u = b_2
\tag{10}
\end{equation}
The acceleration-feedback terms therefore modify a mass-like closed-loop matrix rather than the physical plant inertia.

This can be written as
\begin{equation}
A=\begin{bmatrix} a_{11} & a_{12}\\ a_{21} & a_{22}\end{bmatrix}
\tag{11}
\end{equation}
\begin{align}
a_{11}&=m_s+D_{\mathrm{sky}}, & a_{12}&=D_{\mathrm{gr}}, \tag{11a}\\
a_{21}&=-D_{\mathrm{sky}}, & a_{22}&=m_u-D_{\mathrm{gr}}. \tag{11b}
\end{align}
\begin{equation}
B=\begin{bmatrix} b_1 \\ b_2 \end{bmatrix},
\tag{12}
\end{equation}
\begin{align}
b_1&=-k_s(z_s-z_u)-c_s(\dot z_s-\dot z_u)\\
&-P_{\mathrm{sky}}\dot z_s-P_{\mathrm{gr}}\dot z_u, \tag{12a}\\
b_2&=\phantom{-}k_s(z_s-z_u)+c_s(\dot z_s-\dot z_u)-k_t(z_u-z_r)\\
&+P_{\mathrm{sky}}\dot z_s+P_{\mathrm{gr}}\dot z_u. \tag{12b}
\end{align}
with $[\ddot z_s,\ddot z_u]^\top=A^{-1}B$. At each simulation step, this is a direct $2\times2$ linear solve for $(\ddot z_s,\ddot z_u)$, not a quasi-static approximation.

\subsection{MR Damper Model Integration}
To realize $F_d$, we use the Bouc-Wen MR model
\begin{equation}
F_{\mathrm{MR}}(u)=C(u)\dot z_{\mathrm{rel}}+k_0 z_{\mathrm{rel}}+\alpha(u)\xi
\tag{13}
\end{equation}
with
\begin{align}
C(u)&=C_{0a}+C_{0b}u, \tag{14a}\\
\alpha(u)&=\alpha_a+\alpha_bu. \tag{14b}
\end{align}
Setting $F_{\mathrm{MR}}=F_d$ gives
\begin{equation}
F_d=(C_{0a}+C_{0b}u)\dot z_{\mathrm{rel}}+k_0 z_{\mathrm{rel}}+(\alpha_a+\alpha_bu)\xi
\tag{15}
\end{equation}
\begin{equation}
F_d-C_{0a}\dot z_{\mathrm{rel}}-k_0z_{\mathrm{rel}}-\alpha_a \xi
=u(\alpha_b\xi+C_{0b}\dot z_{\mathrm{rel}})
\tag{16}
\end{equation}
\begin{equation}
u_{\mathrm{req}}=
\frac{F_d-C_{0a}\dot z_{\mathrm{rel}}-k_0z_{\mathrm{rel}}-\alpha_a \xi}
{C_{0b}\dot z_{\mathrm{rel}}+\alpha_b \xi}.
\tag{17}
\end{equation}
\begin{equation}
u=
\begin{cases}
\mathrm{sat}(u_{\mathrm{req}},0,u_{\max}), & |C_{0b}\dot z_{\mathrm{rel}}+\alpha_b \xi|\ge \epsilon,\\
0, & |C_{0b}\dot z_{\mathrm{rel}}+\alpha_b \xi|<\epsilon,
\end{cases}
\tag{18}
\end{equation}
where $\epsilon=10^{-8}$, $u_{\max}=5~\mathrm{V}$, and $\mathrm{sat}(v,0,u_{\max})=\min(\max(v,0),u_{\max})$. If the requested force lies outside the MR operating range, the realized semi-active actuator applies the closest feasible bounded force permitted by $u\in[0,u_{\max}]$ rather than an arbitrary active force.
\begin{figure}[!t]
\centering
\includegraphics[width=\columnwidth]{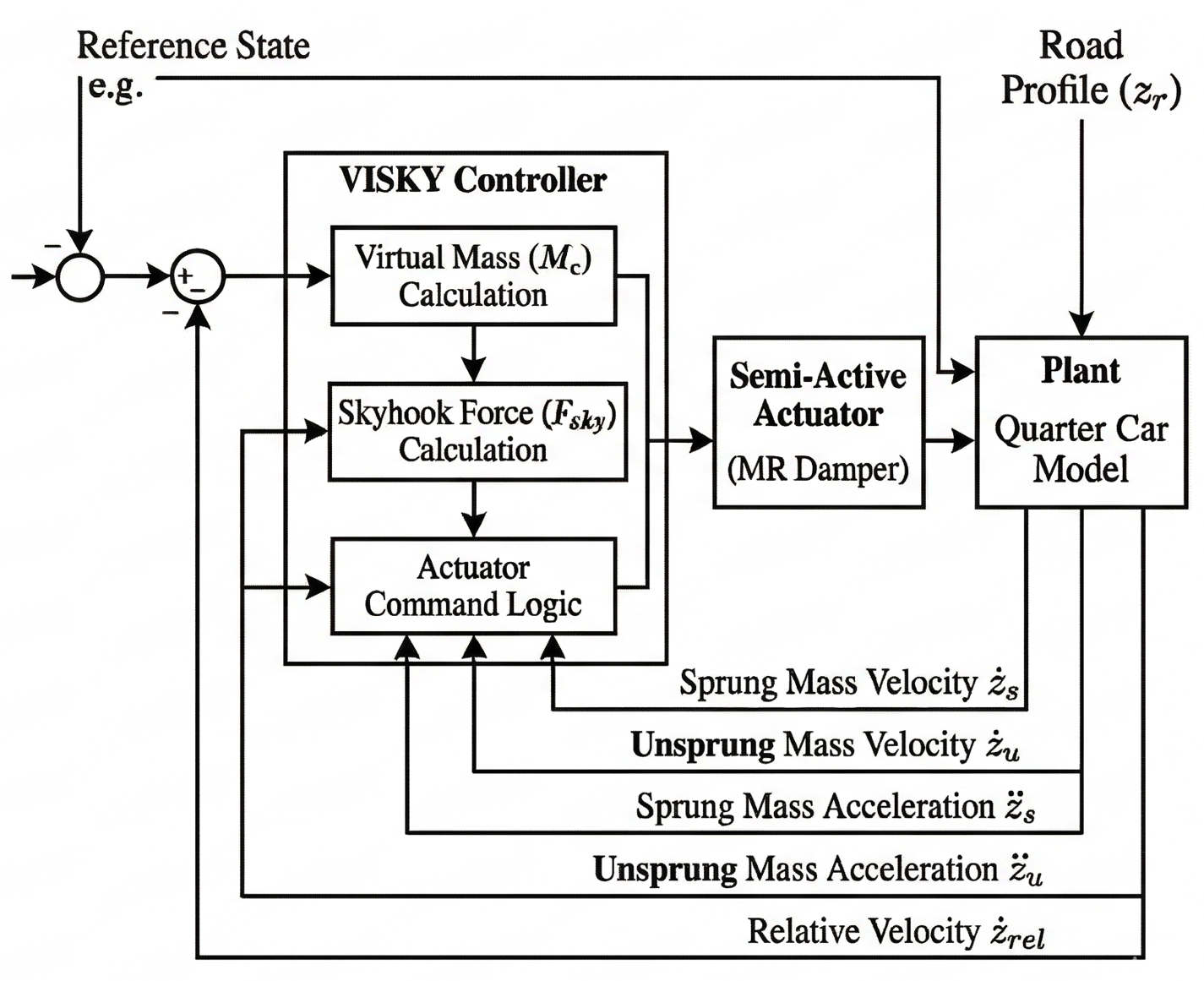}
\caption{VISKY control architecture. Sensor measurements are processed by the VISKY controller to compute the desired damping force, then MR inversion and saturation generate the command applied to the damper in the quarter-car plant.}
\label{fig:control_block}
\end{figure}

\section{Stability Analysis}
Under ideal force tracking $F=F_d$ and constant road input $z_r(t)=z_{r0}$, define shifted coordinates
\begin{equation}
\tilde z_s=z_s-z_{r0},\qquad \tilde z_u=z_u-z_{r0},\qquad
\tilde{\mathbf z}=[\tilde z_s,\tilde z_u]^\top.
\tag{19}
\end{equation}
The equilibrium is $\tilde{\mathbf z}=0$, $\dot{\tilde{\mathbf z}}=0$.

The closed-loop dynamics become
\begin{equation}
M_c\ddot{\tilde{\mathbf z}}+C_c\dot{\tilde{\mathbf z}}+K_c\tilde{\mathbf z}=0
\tag{20}
\end{equation}
with
\begin{equation}
\begin{aligned}
M_c&=
\begin{bmatrix}
m_s+D_{\mathrm{sky}} & D_{\mathrm{gr}}\\
-D_{\mathrm{sky}} & m_u-D_{\mathrm{gr}}
\end{bmatrix},
\\
C_c&=
\begin{bmatrix}
c_s+P_{\mathrm{sky}} & -c_s+P_{\mathrm{gr}}\\
-(c_s+P_{\mathrm{sky}}) & c_s-P_{\mathrm{gr}}
\end{bmatrix}
\end{aligned}
\tag{21}
\end{equation}
\begin{equation}
K_c=
\begin{bmatrix}
k_s & -k_s\\
-k_s & k_s+k_t
\end{bmatrix}.
\tag{22}
\end{equation}
Here, $M_c$ is a mass-like closed-loop matrix induced by acceleration feedback; it is not the physical inertia matrix of the quarter-car.

The characteristic polynomial is
\begin{equation}
\begin{aligned}
p(s)&=\det\!\left(M_c s^2 + C_c s + K_c\right)\\
&=a_4s^4+a_3s^3+a_2s^2+a_1s+a_0
\end{aligned}
\tag{23}
\end{equation}
with coefficients
\begin{align}
a_4&=m_u(m_s+D_{\mathrm{sky}})-m_sD_{\mathrm{gr}}, \tag{23a}\\
a_3&=c_s(m_s+m_u)+m_uP_{\mathrm{sky}}-m_sP_{\mathrm{gr}}, \tag{23b}\\
a_2&=k_s(m_s+m_u)+k_t(m_s+D_{\mathrm{sky}}), \tag{23c}\\
a_1&=k_t(P_{\mathrm{sky}}+c_s), \tag{23d}\\
a_0&=k_sk_t. \tag{23e}
\end{align}

For the quartic polynomial in (23), Routh--Hurwitz conditions are
\begin{align}
a_i&>0,\quad i=0,\dots,4, \tag{24a}\\
a_3a_2&>a_4a_1, \tag{24b}\\
a_3a_2a_1&>a_4a_1^2+a_3^2a_0. \tag{24c}
\end{align}

\noindent\textbf{Proposition 1.} Under ideal force tracking $F=F_d$ and constant road input, the linear closed-loop system in (20) is exponentially stable if (24a)--(24c) hold.

\noindent\emph{Proof.} For the LTI system in (20), $p(s)$ in (23) is the characteristic polynomial. The Routh--Hurwitz conditions in (24a)--(24c) imply that all roots of $p(s)$ lie in the open left half-plane. Therefore all modal components decay exponentially, and the equilibrium of the linear shifted system is exponentially stable. \hfill$\square$

A bounded inversion or saturation error in the practical MR implementation perturbs the ideal force-tracking assumption; Proposition 1 does not by itself prove exponential stability of the full Bouc--Wen plus saturation model.

\section{VISKY Gain Tuning}
\subsection{Admissible Gain Set}
VISKY and baseline gains were tuned by constrained optimization under the same simulation settings and the same acceleration-based objective. The MR command was bounded by $0\le u\le 5$, which corresponds to $C_{\min}=2100\,\mathrm{N\cdot s/m}$ and $C_{\max}=19600\,\mathrm{N\cdot s/m}$ in the Bouc--Wen viscous term.

Each candidate $\theta=[P_{\mathrm{sky}},D_{\mathrm{sky}},P_{\mathrm{gr}},D_{\mathrm{gr}}]^\top$ was required to satisfy:
\begin{enumerate}
\item Invertible mass-like closed-loop matrix: $\det(M_c)=a_4\neq 0$.
\item Routh--Hurwitz inequalities in (24a)--(24c): $a_i>0$ ($i=0,\dots,4$), $a_3a_2>a_4a_1$, and $a_3a_2a_1>a_4a_1^2+a_3^2a_0$.
\end{enumerate}
For the final gains in Table~\ref{tab:tuned_gains}, all admissibility checks are satisfied.

\subsection{Objective and Road Generation}
The optimization objective is
\begin{align}
J(\theta)&=0.5\,\mathrm{RMS}(\ddot z_s)+0.5\,\mathrm{RMS}(\ddot z_u) \tag{25a}\\
\theta&=[P_{\mathrm{sky}},D_{\mathrm{sky}},P_{\mathrm{gr}},D_{\mathrm{gr}}]^\top. \tag{25b}
\end{align}
where $\mathrm{RMS}(\ddot z_s)$ and $\mathrm{RMS}(\ddot z_u)$ are body-acceleration and wheel-acceleration metrics, respectively. The objective therefore targets acceleration response only; suspension travel and dynamic tire load are treated as secondary outcomes in Section VI. Fine tuning was performed on ISO 8608-consistent random roads generated from the spatial PSD model
\begin{align}
G_q(n)&=G_q(n_0)\left(\frac{n}{n_0}\right)^{-w},\quad n_0=0.1~\mathrm{m^{-1}},\;w=2
\tag{26a}\\
z_r(s)&=\sum_{i=1}^{N}\sqrt{2\,G_q(n_i)\,\Delta n}\cos\!\left(2\pi n_i s+\phi_i\right),
\tag{26b}
\end{align}
where $\phi_i\sim\mathcal{U}(0,2\pi)$ and the spatial coordinate satisfies $s=vt$ for vehicle speed $v$. In the reported tuning setup, ISO classes B, C, and D were synthesized over a 100~m road length with spatial sampling $\Delta s=0.02~\mathrm{m}$, retained spatial-frequency band $0.01\le n\le 10~\mathrm{m^{-1}}$, and $N=1000$ retained spatial-frequency bins. Tuning scenarios were evaluated at $v\in\{10,20\}~\mathrm{m/s}$, with class-dependent random phases generated from seed 42 and deterministic class offsets.

\begin{figure}[!htbp]
\centering
\includegraphics[width=\columnwidth]{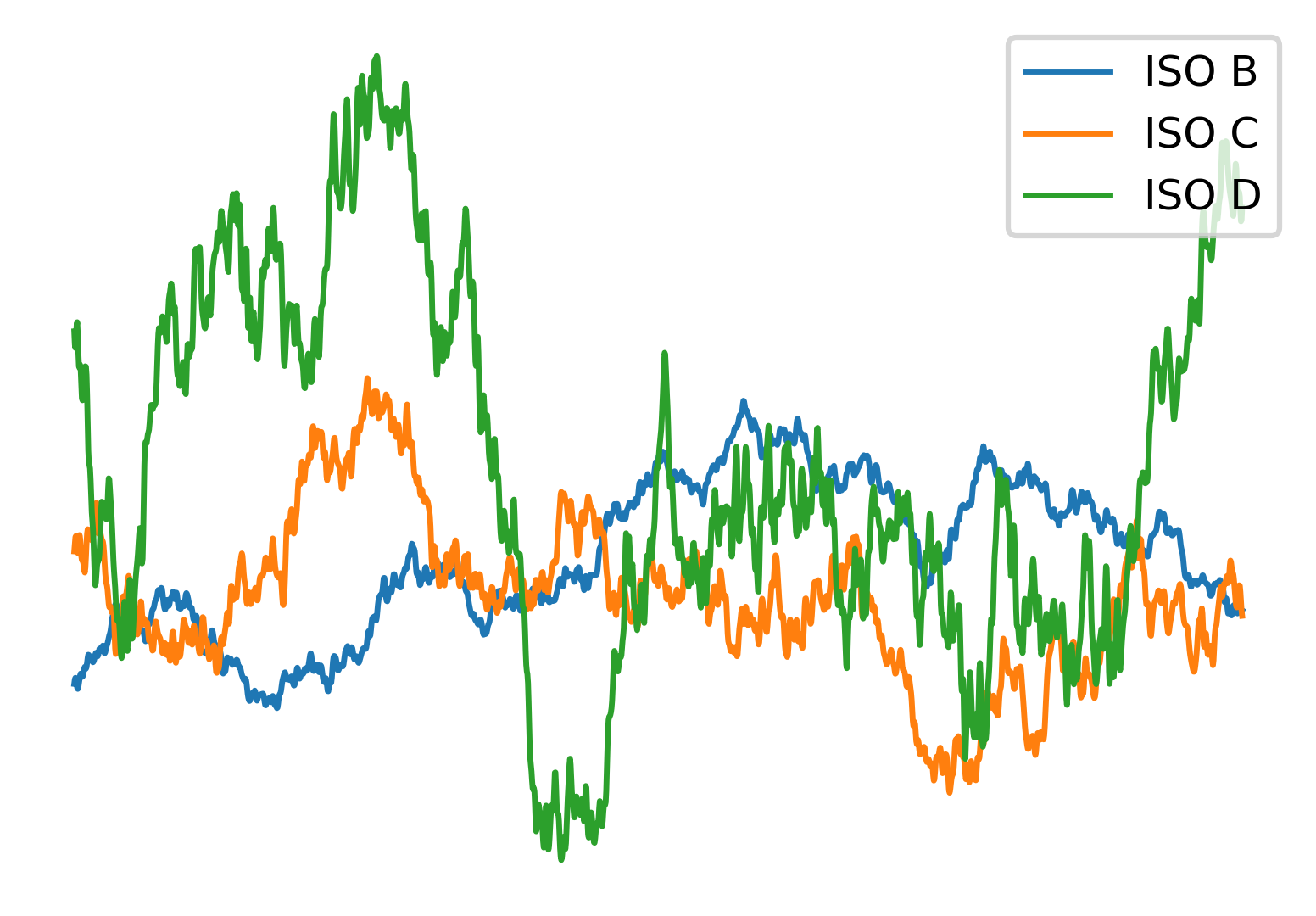}
\caption{Representative ISO 8608 spatial road profiles (classes B, C, and D) used for gain tuning.}
\label{fig:random_road}
\end{figure}

Representative ISO 8608 road profiles used in tuning are shown in Fig.~\ref{fig:random_road}.

\subsection{Search Procedure and Tuned Gains}
The reported tuning did not use analytical gradient descent. Instead, it used a stability-constrained sample-based search:
\begin{enumerate}
\item Draw global random candidate gains inside predefined bounds.
\item Reject candidates violating $\det(M_c)\neq 0$ or (24a)--(24c).
\item Refine the best feasible candidates with local stochastic perturbations.
\end{enumerate}
Each controller was tuned separately using the same ISO road scenarios, the same acceleration-based objective, and the same global/local search budget. This ensures that the comparison between Skyhook, Groundhook, Skygroundhook, and VISKY is fair at the level of tuning procedure.

Table~\ref{tab:tuned_gains} summarizes the tuned gains for VISKY and baseline controllers.
These entries are controller-law gains used to compute the desired semi-active force; they are not direct physical MR damping coefficients. The realized MR damper remains bounded by $u\in[0,u_{\max}]$ and therefore by $C(u)\in[C_{\min},C_{\max}]$ through the inversion and saturation in (18).
\begin{table}[!htbp]
\caption{Tuned Controller Gains}
\label{tab:tuned_gains}
\centering
\scriptsize
\resizebox{\columnwidth}{!}{%
\begin{tabular}{l p{0.40\columnwidth} p{0.30\columnwidth}}
\toprule
Method & Parameter(s) & Value(s) \\
\midrule
Skyhook & $C_{\mathrm{sky}}$ & 17000 \\
Groundhook & $C_{\mathrm{gr}}$ & 4000 \\
Skygroundhook & $C_{\mathrm{sky}},C_{\mathrm{gr}}$ & $25500,1150$ \\
VISKY & $P_{\mathrm{sky}},D_{\mathrm{sky}},P_{\mathrm{gr}},D_{\mathrm{gr}}$ & $7400,5600,440,50$ \\
\bottomrule
\end{tabular}
}
\end{table}

Table~\ref{tab:model_params} lists the quarter-car and Bouc--Wen MR damper parameters used in simulation.
Quarter-car parameter values were selected to be consistent with representative values reported in \cite{c5,c9,c10,c25}, and the MR damper (Bouc--Wen) parameters were taken from \cite{c26}.
\begin{table}[!htbp]
\caption{Model Parameters Used in Simulation}
\label{tab:model_params}
\centering
\scriptsize
\begin{tabular}{l p{0.37\columnwidth} p{0.30\columnwidth}}
\toprule
Parameter & Description & Value \\
\midrule
\multicolumn{3}{c}{\textbf{Quarter-Car Parameters}} \\
\midrule
$m_s$ & Sprung mass & $320\,\mathrm{kg}$ \\
$m_u$ & Unsprung mass & $45\,\mathrm{kg}$ \\
$c_s$ & Passive damping coefficient & $1500\,\mathrm{N\cdot s/m}$ \\
$k_s$ & Suspension spring constant & $22000\,\mathrm{N/m}$ \\
$k_t$ & Tire stiffness & $192000\,\mathrm{N/m}$ \\
\midrule
\multicolumn{3}{c}{\textbf{Bouc--Wen MR Damper Parameters}} \\
\midrule
$C_{0a}$ & Base viscous coefficient & $2100\,\mathrm{N\cdot s/m}$ \\
$C_{0b}$ & Viscous slope wrt. voltage & $3500\,\mathrm{N\,s\,m^{-1}\,V^{-1}}$ \\
$\alpha_a$ & Base elastic term & $1400\,\mathrm{N}$ \\
$\alpha_b$ & Elastic slope wrt. voltage command & $69500\,\mathrm{N/V}$ \\
$k_0$ & Elastic stiffness coefficient & $0\,\mathrm{N/m}$ \\
$q$ & Bouc--Wen parameter 1 & 48000 \\
$b$ & Bouc--Wen parameter 2 & 48000 \\
$\gamma$ & Bouc--Wen parameter 3 & 4 \\
$u_{\max}$ & Maximum damper input voltage & $5\,\mathrm{V}$ \\
\bottomrule
\end{tabular}
\end{table}

\FloatBarrier
\section{Results and Discussion}
\subsection{Simulation Results}
\begin{table*}[!t]
\caption{Bump-Road Test Results}
\label{tab:perf_bump}
\centering
\scriptsize
\resizebox{\textwidth}{!}{%
\begin{tabular}{lcccc}
\toprule
Method & RMS Sprung-Mass Acceleration ($\mathrm{m/s^2}$) & RMS Unsprung-Mass Acceleration ($\mathrm{m/s^2}$) & RMS Suspension Travel (mm) & RMS Dynamic Tire Load (N) \\
\midrule
Passive MR & 0.086711 & 0.060631 & 6.6686 & 187.98 \\
Skyhook & 0.058544 (+32.484\%) & 0.061351 (-1.188\%) & 3.9328 (+41.025\%) & 152.52 (+18.864\%) \\
Groundhook & 0.084340 (+2.734\%) & 0.061527 (-1.478\%) & 5.4033 (+18.974\%) & 186.54 (+0.766\%) \\
Skygroundhook & 0.058004 (+33.107\%) & 0.061401 (-1.270\%) & 3.3977 (+49.049\%) & 161.90 (+13.874\%) \\
VISKY & \textbf{0.045467 (+47.565\%)} & \textbf{0.059609 (+1.686\%)} & \textbf{3.9751 (+40.391\%)} & \textbf{108.55 (+42.254\%)} \\
\bottomrule
\end{tabular}
}
\end{table*}

\begin{table*}[!t]
\caption{ISO 8608 Random-Road Test Results}
\label{tab:iso8608_perf}
\centering
\scriptsize
\resizebox{\textwidth}{!}{%
\begin{tabular}{lcccc}
\toprule
Method & RMS Sprung-Mass Acceleration ($\mathrm{m/s^2}$) & RMS Unsprung-Mass Acceleration ($\mathrm{m/s^2}$) & RMS Suspension Travel (mm) & RMS Dynamic Tire Load (N) \\
\midrule
Passive MR & 0.05098 & 0.12220 & 5.358 & 418.61 \\
Skygroundhook & 0.031533 (+38.146\%) & 0.101804 (+16.691\%) & 5.010146 (+6.4922\%) & 434.15 (-3.7123\%) \\
\textbf{VISKY} & \textbf{0.030002 (+41.158\%)} & \textbf{0.099854 (+18.288\%)} & \textbf{4.8956 (+8.6416\%)} & \textbf{432.23 (-3.2509\%)} \\
\bottomrule
\end{tabular}
}
\end{table*}

The passive quarter-car natural frequencies are obtained from
\begin{equation}
(-\omega^2M+K)\mathbf Q=0,\quad \mathbf q=\mathbf Qe^{j\omega t}
\tag{27}
\end{equation}
with characteristic equation
\begin{equation}
\det(-\omega^2M+K)=0,
\tag{27a}
\end{equation}
\begin{equation}
\begin{vmatrix}
k_s-\omega^2m_s & -k_s\\
-k_s & k_s+k_t-\omega^2m_u
\end{vmatrix}=0,
\tag{27b}
\end{equation}
\begin{equation}
\omega^4m_sm_u-\omega^2(k_sm_u+k_sm_s+k_tm_s)+k_sk_t=0,
\tag{27c}
\end{equation}
which yields $\omega_1=7.85~\mathrm{rad/s}$ and $\omega_2=69.01~\mathrm{rad/s}$. These correspond to the passive nominal body-bounce mode ($1.25~\mathrm{Hz}$) and wheel-hop mode ($10.98~\mathrm{Hz}$), respectively. They are used as reference frequencies for comparison; they are not controller-dependent peak locations.

\begin{table}[!htbp]
\caption{Stepped Sine-Sweep Results}
\label{tab:sine}
\centering
\small
\resizebox{\columnwidth}{!}{%
\begin{tabular}{ccc}
\toprule
Method & $|Z_s/Z_r|$ at $\omega_1=7.85~\mathrm{rad/s}$ & $|Z_s/Z_r|$ at $\omega_2=69.01~\mathrm{rad/s}$ \\
\midrule
Passive MR & 18.958 & 8.7626 \\
Skyhook & 7.9334 (+58.154\%) & 9.1375 (-4.2789\%) \\
Groundhook & 18.578 (+2.008\%) & 8.5259 (+2.7014\%) \\
Skygroundhook & 9.0566 (+52.229\%) & 8.7405 (+0.25146\%) \\
VISKY & \textbf{8.2235 (+56.623\%)} & \textbf{2.2850 (+73.923\%)} \\
\bottomrule
\end{tabular}
}
\end{table}

\begin{figure}[H]
\centering
\includegraphics[width=\columnwidth]{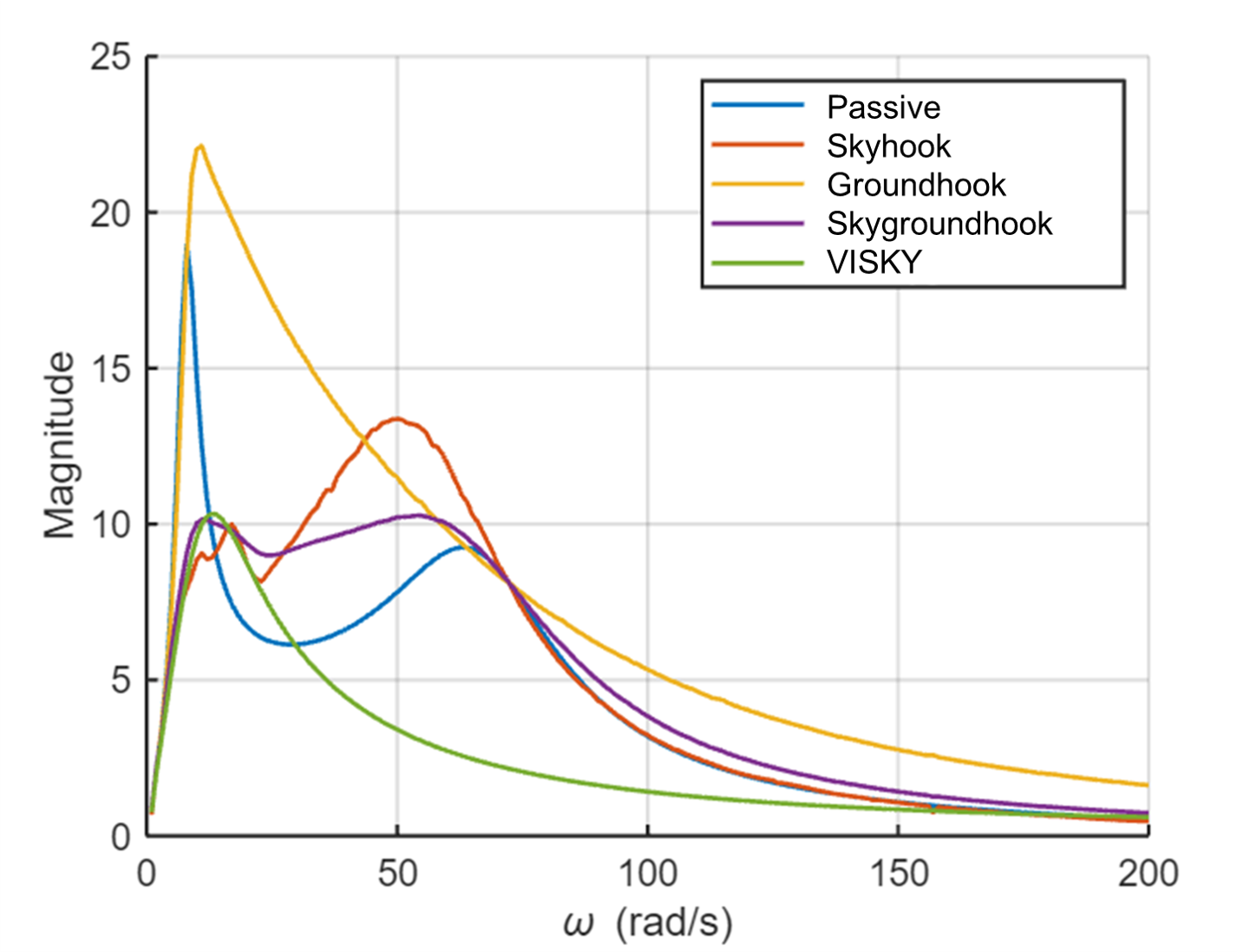}
\caption{Stepped sine-sweep comparison. The plotted magnitude is the sprung-displacement transmissibility $|Z_s|_{\mathrm{RMS}}/|Z_r|_{\mathrm{RMS}}$ versus excitation frequency. VISKY nearly preserves Skyhook's first-mode attenuation while strongly reducing the second-mode wheel-hop response.}
\label{fig:sine_compare}
\end{figure}

A stepped sine sweep was then applied with road amplitude $5~\mathrm{mm}$ over 18 logarithmically spaced frequencies from 4 to $90~\mathrm{rad/s}$. Because the test is stepped rather than continuous, no single sweep rate applies; at each frequency, four cycles were discarded and the next two cycles were used for RMS evaluation. Table~\ref{tab:sine} and Fig.~\ref{fig:sine_compare} summarize the comparison. In Table~\ref{tab:sine}, the reported values are modal displacement transmissibility magnitudes at the passive nominal modes; lower values are better, and the percentages are relative to the passive MR baseline. At $\omega_1$, pure Skyhook gives the lowest transmissibility, but VISKY remains close and is 9.2\% lower than Skygroundhook. At $\omega_2$, VISKY reduces the peak by 73.9\% relative to Skygroundhook, showing that the added acceleration feedback is most beneficial in the wheel-hop range.

The bump road input is
\begin{equation}
z_r(t)=\begin{cases}
\dfrac{1}{2}h_b\left(1-\cos\left(2\pi\,\dfrac{t-T_1}{T_2-T_1}\right)\right), & T_1\le t\le T_2,\\
0, & \text{otherwise}.
\end{cases}
\tag{28}
\end{equation}

The bump-excitation results are summarized in Table~\ref{tab:perf_bump}. All quantities are RMS metrics; lower values are better, and the percentages are relative to the passive MR baseline. Table~\ref{tab:perf_bump} shows that VISKY reduces RMS sprung-mass acceleration by 47.6\% relative to passive MR damping. Relative to Skygroundhook, VISKY lowers sprung-mass acceleration by 21.6\%, unsprung-mass acceleration by 2.9\%, and dynamic tire load by 33.0\%, but it increases suspension travel by 17.0\%. This trade-off is consistent with the tuning objective in Section V, which targets acceleration metrics rather than travel directly.

Table~\ref{tab:iso8608_perf} summarizes the representative ISO 8608 random-road test. The table reports RMS metrics; lower values are better, and the percentages are relative to the passive baseline in the first row. Relative to Skygroundhook, VISKY lowers sprung-mass acceleration by 4.9\%, unsprung-mass acceleration by 1.9\%, suspension travel by 2.3\%, and dynamic tire load by 0.44\%. The smaller margin, compared with the stepped sine-sweep wheel-hop case, is expected under broadband excitation because the input energy is distributed across frequencies instead of being concentrated near a single resonance.

Additional ISO 8608 time- and frequency-domain responses are provided in Appendix~A.

\FloatBarrier
\subsection{Evaluation}
VISKY nearly preserves the first-mode body-bounce attenuation of pure Skyhook while substantially reducing wheel-hop response relative to Skygroundhook. This behavior is physically consistent with the control law: the acceleration-feedback terms become more influential as frequency increases, so the associated virtual inertia matrix is most effective in the wheel-hop band. The results also show that VISKY is not uniformly dominant in every secondary metric. In particular, the bump test shows a suspension-travel penalty relative to Skygroundhook even though the acceleration and tire-load metrics improve. That trade-off is expected because the tuning objective uses acceleration metrics only.
\FloatBarrier
\section{Conclusion}
This paper presented VISKY, a sky-ground hybrid semi-active controller with additional acceleration feedback for MR dampers. In the closed-loop quarter-car model, these acceleration terms appear as a virtual inertia matrix, which provides a compact interpretation of the controller's improved high-frequency behavior. In the reported simulations, VISKY preserved first-mode comfort close to Skyhook, improved acceleration metrics relative to Skygroundhook under bump and representative ISO 8608 random-road inputs, and strongly reduced the wheel-hop peak in stepped sine sweep. The main trade-off is that acceleration-focused tuning does not guarantee the smallest suspension travel in every case. The controller remains attractive for embedded implementation because it requires only a small linear solve and bounded MR inversion.

\clearpage
\onecolumn
\useRomanappendicesfalse
\appendices
\section{Additional ISO 8608 Responses}
This appendix provides representative time- and frequency-domain responses for the ISO 8608 random-road test summarized in Table~\ref{tab:iso8608_perf}. The panels separate acceleration, suspension-travel, tire-load, frequency-domain, and auxiliary MR-damper signals for direct inspection.
\begin{figure}[H]
\centering
\begin{minipage}[t]{0.48\textwidth}
\centering
\includegraphics[width=\linewidth]{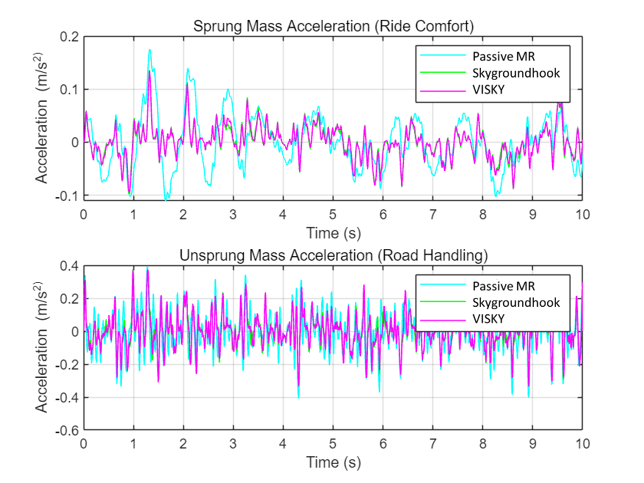}
\end{minipage}\hfill
\begin{minipage}[t]{0.48\textwidth}
\centering
\includegraphics[width=\linewidth]{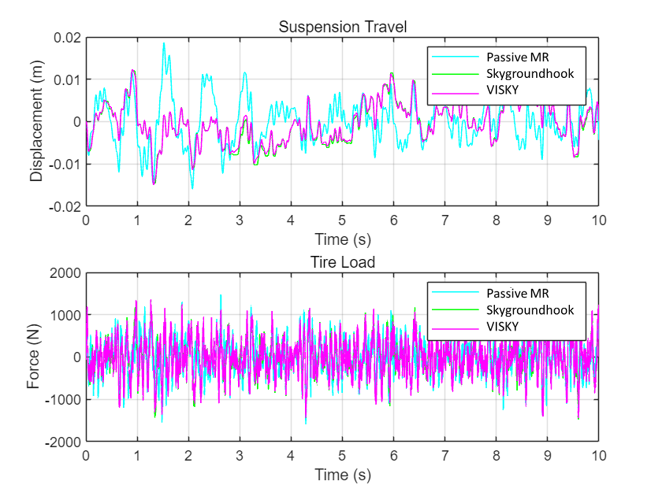}
\end{minipage}

\vspace{1.5mm}

\begin{minipage}[t]{0.48\textwidth}
\centering
\includegraphics[width=\linewidth]{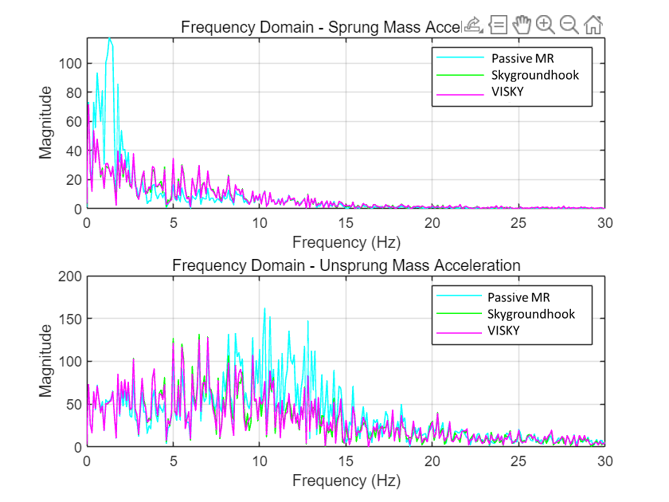}
\end{minipage}\hfill
\begin{minipage}[t]{0.48\textwidth}
\centering
\includegraphics[width=\linewidth]{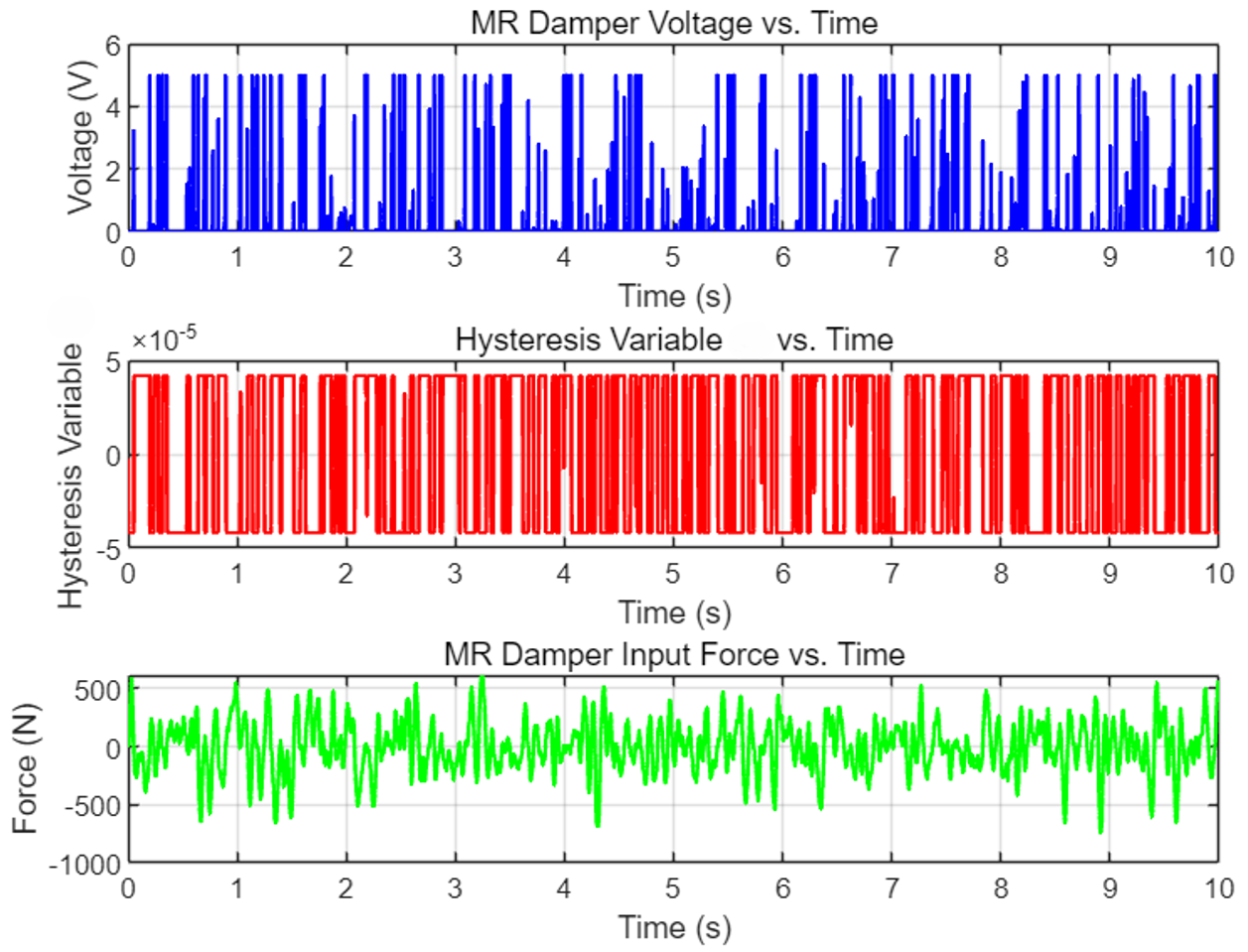}
\end{minipage}
\caption{Representative ISO 8608 random-road responses. The four panels show acceleration time histories, suspension-travel and dynamic-tire-load histories, frequency-domain acceleration responses, and additional MR-damper signals for the same test case summarized in Table~\ref{tab:iso8608_perf}.}
\label{fig:iso8608_appendix_abcd}
\end{figure}

\end{document}